\documentclass[a4paper]{jpconf}
\usepackage{graphicx}
\usepackage{amsmath}
\usepackage{amssymb}
\begin{document}
\title{Pulsating Strings in Deformed Backgrounds}

\author{Victor O. Rivelles}

\address{Instituto de F\'{i}sica, Universidade de S\~ao Paulo, C. Postal 66318, 05314-970 S\~ao Paulo, SP, Brazil}

\ead{rivelles@fma.if.usp.br}

\begin{abstract}
This is a brief summary on pulsating strings in beta deformed backgrounds found recently. 
\end{abstract}

\section{Introduction}

Spinning and rotating strings are well understood in the context of the AdS/CFT correspondence and they have been used to provide strong support to the duality between type IIB strings in $AdS_5 \times S^5$ and ${\cal N}=4$ super Yang-Mills in four dimensions. On the other side  pulsating strings are much less known. They have been studied in $AdS_5 \times S^5$ \cite{deVega:1994yz,Gubser:2002tv,Minahan:2002rc,Engquist:2003rn,Khan:2003sm,Arutyunov:2003za,Kruczenski:2004cn,Smedback:1998yn} but the corresponding operators in the gauge theory are still poorly understood. 

It is also desirable to extend the AdS/CFT correspondence to situations where the number of supersymmetries is less than the maximal allowed. A marginal deformation of the ${\cal N}=4$ theory which preserves ${\cal N}=2$ supersymmetries and depends on one parameter $\beta$ is known for a long time \cite{Leigh:1995ep,Mauri:2005pa}. Its gravitational dual, found by Lunin and Maldacena \cite{Lunin:2005jy}, corresponds to a string propagating in $AdS_5 \times S^5_{\hat{\gamma}}$, a background where the five sphere has a deformation depending on one parameter $\hat{\gamma}$. Rotating and spinning strings have been studied in Lunin-Maldacena backgrounds given further support to the correspondence. We will present some results for pulsating strings in this deformed background which were obtained in collaboration with S. Giardino \cite{Giardino:2011jy}. 

Since the string has a periodic motion we will use the string oscillation number and its  energy to characterize its classical behavior. Notice that the string oscillation number is not one of the string charges but being an adiabatic invariant it provides information about the high energy regime. We will consider the case of highly excited string states. We first consider pulsating strings in deformed Minkowski spacetime. After quantization we look for highly excited string states for small deformation and compute its energy in perturbation theory. We find full agreement with the quantum results if the oscillation number is quantized as an even number. In the $AdS_5 \times S^5_{\hat{\gamma}}$ case we consider only strings oscillating in the deformed sphere since $AdS_5$ is not deformed. We find that the oscillation number can be expressed in terms of elliptic functions. We derive the relation between the energy and the oscillation number for short strings in the low energy regime. When the deformation vanishes we recover the results for $AdS_5 \times S^5$ \cite{Beccaria:2010zn}. In the case of large energy we consider the quantization of highly excited strings up to second order in perturbation theory and we find that the energy has a new term proportional to the deformation parameter which is not present in the classical relation between the energy and the oscillation number. We also discuss the contributions coming from the fluctuations of the radial $AdS$ coordinate. 

It is known that the gauge theory operator dual to pulsating strings in $AdS_5 \times S^5$ is composed of non holomorphic products of the complex scalar fields and that there is a precise match between the energy and the anomalous dimension \cite{Engquist:2003rn,Kruczenski:2004cn,Beisert:2003xu,Minahan:2004ds}. It would be interesting to extend these results to the deformed case using the results of \cite{Gromov:2010dy}. Notice that for a rotating string in $AdS_5 \times S^5_{\hat{\gamma}}$ with angular momentum $J$ and winding number $m$ it is known that its energy can be obtained from the $AdS_5 \times S^5$ energy simply by replacing $|m|$ by $|m + \frac{1}{2} \hat{\gamma} J|$ \cite{Frolov:2005ty}. This is also true for the fluctuations around the classical solution. For pulsating strings such a property no longer holds.

\section{Pulsating Strings in Deformed Flat Spacetime}

The ten dimensional Minkowski spacetime is split into a four dimensional one and the remaining six dimensional space is deformed resulting in the deformed flat spacetime given by \cite{Lunin:2005jy} 
\begin{eqnarray}
ds^{2}&=&\eta_{\mu\nu}dx^{\mu}dx^{\nu}+\sum_{i=1}^{3}\left(dr_{i}^{2}+Gr_{i}^{2}d\phi_{i}^{2}\right)+\gamma^{2}r_{1}^{2}r_{2}^{2}r_{3}^{2}G\left(\sum_{i=1}^{3}d\phi_{i}\right)^{2}, \label{R10
deformado}\\
B_{2}&=& \gamma 
G\left(r_{1}^{2}r_{2}^{2}\, d\phi_{1}\wedge
  d\phi_{2}+r_{2}^{2}r_{3}^{2}\, d\phi_{2}\wedge
  d\phi_{3}-r_{1}^{2}r_{3}^{2}\, d\phi_{1}\wedge
  d\phi_{3}\right), \nonumber\\
  e^{2\Phi}&=&G,\qquad
G^{-1}=1+\gamma^{2}\left(r_1^2r_2^2+r_2^2r_3^2+r_1^2r_3^2\right).\nonumber
\end{eqnarray}
No RR fields are present in this case. As expected, when the deformation vanishes, $\gamma=0$, we get ten dimensional Minkowski spacetime. 

\begin{figure}[h] 
       \centering  
       \includegraphics[height=4cm]{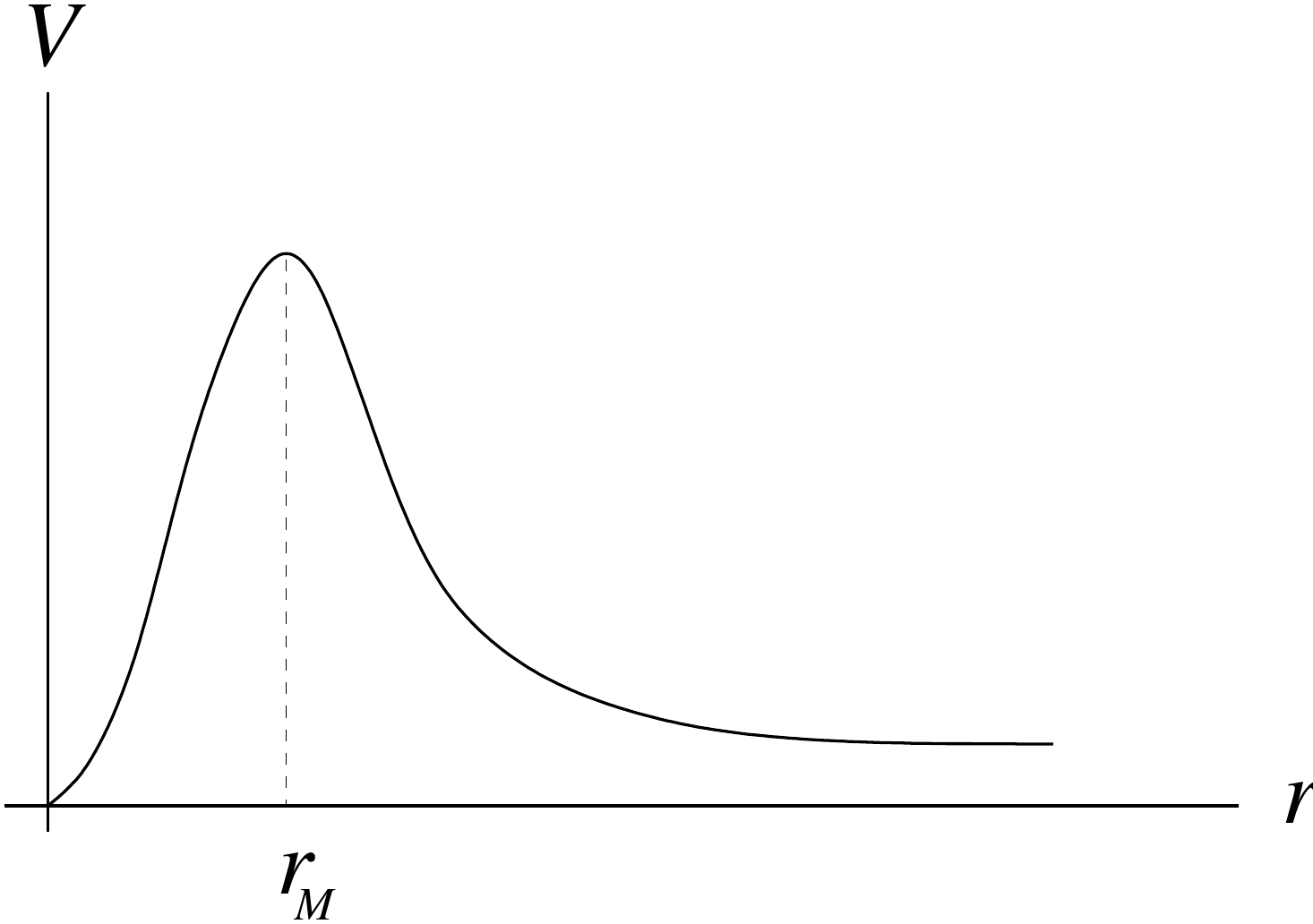}
       \caption{\it The effective potential in deformed Minkowski spacetime}\label{figA}
\end{figure}

We now consider a pulsating strings in (\ref{R10 deformado}). We will use spherical coordinates defined by
\begin{equation}
	r_1 = r \sin\theta\cos\psi,  \qquad r_2 = r \sin\theta \sin\psi, \qquad r_3 = r \cos\theta, 
\end{equation}
and choose the string at the origin of the Minkowski spacetime with  $\psi=\pi/2$, $\phi_1 = \phi_{2}=0$ and  $\theta$ fixed so that the relevant part of the metric becomes
\begin{eqnarray}
ds^2&=&-dt^2+dr^2 + G r^2 \cos^2\theta d\phi_3^2, \label{metr1}\\
G^{-1}&=&1+\gamma^{2}r^{4}\sin^2\theta\cos^2\theta.\label{(G-1)}
\end{eqnarray}
The coupling of the string to the $B_2$ field vanishes. The ansatz for a pulsating string that is wound in $\phi_3$ and oscillating in the radial direction is  
\begin{equation}
t=\kappa\tau, \qquad r=r(\tau), \qquad \phi_3=m\sigma, \qquad \theta = \mbox{constant},
\label{ansatz minahan R}
\end{equation}
with $m$ being the winding number.  
After the rescale $m$ and $\gamma$ as  $m \cos\theta \rightarrow m$ and $ \gamma \sin\theta \cos \theta \rightarrow \gamma$ respectively, assuming that $\theta$ is different from zero and $\pi/2$, we can rewrite the only non-trivial equation of motion in terms of an effective potential as 
\begin{equation}
\frac{\dot{r}^2}{m^2} = \frac{\kappa^2}{m^2} - V(r), \label{r ponto}
\end{equation}
where $V(r)$ is given by
\begin{equation}
V(r)=\frac{ r^2}{1+\gamma^{2}r^{4}},
\label{V(r)}
\end{equation}
and plotted in Fig. \ref{figA}. There $r_{M}^2=1/\gamma$ is the point where the potential reaches is maximum $V(r_{M})=1/(2\gamma)$. The condition for oscillatory motion implies that $\kappa^2/m^2 \le 1/(2\gamma)$.
From now on we will only consider the case of small deformation $\gamma<<1$. Then the height of the potential is very large and its position is far away from the origin and we can consider a highly excited pulsating string.

The string dynamics can be characterized by the relation between the string oscillation number and the energy. The oscillation number associated to the periodic radial motion is $N = \oint \Pi_r dr/2\pi$, where $\Pi_r$ is the momentum conjugated to $r$. To compute $\Pi_r$ we take the Polyakov action for the ansatz (\ref{ansatz minahan R})
\begin{equation}
S=-\frac{\sqrt{\lambda} \kappa}{2}\int d\tau \left( 1 - \dot{r}^2 + m^2 r^2 G \right), 
\end{equation} 
and find 
\begin{equation} \label{energy}
	E =  \sqrt{2 \lambda^{1/2} m N} \left( 1 - \frac{5}{8} \frac{\gamma^2}{\lambda} \frac{N^2}{m^2}  \right).
\end{equation}
Since adiabatic invariants can be used to probe the highly excited states of the quantum theory the above expression can be regarded an approximation for the energy in the limit of large quantum numbers in the radial direction.  

To quantize  we will use the Nambu-Goto action for the ansatz (\ref{ansatz minahan R})
\begin{equation}
	S = - m \frac{\sqrt{\lambda}}{\kappa} \int dt \,  r \, \sqrt{G} \sqrt{1 - \dot{r}^2}, 
\end{equation}
finding the Hamiltonian 
\begin{equation}
H^2 = \Pi_r^{2} + \lambda m^2 r^2  - \gamma^2 \lambda m^2 \frac{r^6}{1 + \gamma^2 r^4}.  \label{Hamilt Min} 
\end{equation}
If the deformation vanishes we get a radial harmonic oscillator potential $\lambda m^2 r^2$.
We assume that the wave function depends only on $r$ so we have to realize $\Pi_r^2$ as the radial component of the Laplacian
\begin{equation}\label{lap}
	\Pi_r^2 = - \frac{1}{\sqrt{-g}} \frac{d}{dr} \left( \sqrt{-g} \frac{d}{dr} \right).
\end{equation}
First consider the situation where we quantize only the radial motion on the deformed plane (\ref{metr1}) ignoring the remaining coordinates, that is we take $\sqrt{-g} = r \cos\theta \sqrt{G}$ in (\ref{lap}). The Schr\"{o}dinger equation is then 
\begin{equation}\label{H1}
	H^2 \Psi = - \frac{1}{r} \frac{d}{dr} \left( r \frac{d\Psi}{dr} \right) + \lambda m r^2 \Psi + 2 \gamma^2 r^3 \frac{d\Psi}{dr} - \gamma^2 \lambda m^2 \frac{r^6}{1 + \gamma^2 r^4} \Psi = E^2 \Psi.
\end{equation}
To solve (\ref{H1}) we will use standard perturbation theory and we have to choose the unperturbed Hamiltonian. We can either choose the two first terms or the three first terms. In the last case we are considering the quantization in the deformed space while in the first one we are regarding the deformation as a perturbation and quantizing in the undeformed space. In both ways we get the same result at the end. If we choose as the unperturbed Hamiltonian the first two terms of (\ref{H1}) we find that the normalized wave function is 
\begin{equation}
	\Psi_n(r) = \sqrt{2 \lambda^{1/2} m} \,\, e^{-\frac{1}{2} \sqrt{\lambda} m r^2} L_n(\sqrt{\lambda} m r^2),
\end{equation}
where $L_{n}$ are Laguerre polynomials. For highly excited states we have
\begin{equation}
	E^2_{0,n} = 4\sqrt{\lambda} \, m \, n, \label{energy1}
\end{equation}
or $E_{0,n} = \sqrt{ 2 \lambda^{1/2} m (2n)}$. Since we are quantizing a radial harmonic oscillator we expect that its energy depends only on even integers. Comparison with the lowest order of (\ref{energy}) shows complete agreement if the oscillation number is quantized as $2n$. 

The first order correction to the energy for large $n$ is 
\begin{equation} \label{correct1}
	E_n = \sqrt{2\lambda^{1/2} m (2n)} \left( 1 - \frac{5}{8} \frac{\gamma^2}{\lambda} \frac{(2n)^2}{m^2} \right).
\end{equation}
As expected it agrees with (\ref{energy}) if we assume that the oscillation number $N$ is quantized with eigenvalue $2n$.

Some quantum effects of the deformed ten dimensional space can be seen by taking in  $\Pi_r^2$ the contribution of all dimensions by choosing $\sqrt{-g} = r^5 \sqrt{G} \dots$ where the dots represent terms which do not depend on $r$ and cancel out in (\ref{lap}). We also  assume that the wave function depends only on the radial coordinate. The Schr\"{o}dinger equation for $H^2$ reads now 
\begin{equation}
	H^2 \Psi = - \frac{1}{r^5} \frac{d}{dr} \left( r^5 \frac{d \Psi}{dr} \right) + \lambda m^2 r^2 \Psi + 4 \gamma^2 r^3 \frac{d \Psi}{dr} - \gamma^2 \lambda m^2 \frac{r^6}{1 + \gamma^2 r^4} \Psi = E^2 \Psi. \label{ham_flat}
\end{equation}
The normalized wave function is now
\begin{equation}
	\Psi_n(r) = \sqrt{2} \lambda^{3/4} \frac{m^{3/2}}{n} e^{-\sqrt{\lambda} m r^2/2} L^{(2)}_n(\sqrt{\lambda} m r^2),
\end{equation}
where $L^{(2)}_n$ are generalized Laguerre polynomials. The energy for high quantum number is
still given by the former result (\ref{energy1}). This happens because only the zero point contribution to the energy changes with the dimension and in the limit of large $n$ it can be ignored. The corrections coming from the deformation lead to the previous result (\ref{correct1}). This shows that for a highly excited string the classical result (\ref{energy}) can be straightforwardly taken to quantum case when $N$ is quantized as $2n$.

\section{Pulsating Strings in the Deformed Sphere of $AdS_5 \times S^5_{\hat{\gamma}}$}

The the Lunin-Maldacena background is given by \cite{Lunin:2005jy}
\begin{eqnarray}
ds^{2}&=&R^{2}\left[ds_{AdS_{5}}^{2}+\sum_{i=1}^{3}(d\mu_{i}^{2}+G\mu_{i}^{2}d\phi_{i}^2)+\hat{\gamma}^{2}\mu_{1}^{2}\mu_{2}^{2}\mu_{3}^{2}{G\left(\sum_{i=1}^{3}d\phi_{i}\right)}^{2}\right], \label{AdS5 x S5 def}\\
B_{2}&=&\hat{\gamma}R^{2}G\,\left(\mu_{1}^{2}\mu_{2}^{2}\,
  d\phi_{1}\wedge d\phi_{2}+\mu_{2}^{2}\mu_{3}^{2}\, d\phi_{2}\wedge
  d\phi_{3}-\mu_{1}^{2}\mu_{3}^{2}\, d\phi_{1}\wedge
  d\phi_{3}\right), \nonumber
\\e^{2\Phi}&=&e^{2\Phi_{0}}G, \nonumber\\
C_{2}&=&-48\pi\, N\,\hat{\gamma}\,\omega_{1}\wedge d\psi, \nonumber \\
C_{4}&=&16\pi\, N\,(\omega_{4}+G\, d\omega_{1}\wedge d\phi_{1}\wedge d\phi_{2}\wedge
d\phi_{3}), \nonumber
\end{eqnarray}
where $B_2$ is the NS-NS two-form potential, $C_2$ and $C_4$ are the two and four-form RR potentials respectively, $\Phi$ is the dilaton, and 
\begin{eqnarray}
&&ds_{AdS_5}^{2}=-\cosh^2\rho\,dt^2+d\rho^2+\sinh^2\rho\left(d\Psi^2+\sin^2\Psi
  d\Phi_1^2+\cos^2\Psi d\Phi_2^2\right),\label{ads_sec}\\
&& G^{-1}=1+\hat{\gamma}^{2}\left(\mu_{1}^{2}\mu_{2}^{2}+\mu_{2}^{2}\mu_{3}^{2}+\mu_{1}^{2}\mu_{3}^{2}\right),\\
&& {d\omega}_{1}=\sin^{3}\alpha\cos\alpha\,\sin\theta\cos\theta
d\alpha\wedge d\theta, \qquad {\omega}_{AdS_{5}}=d\omega_{4}, \\
&& \sum_{i=1}^{3}\mu_{i}^{2}=1,\qquad\hat{\gamma}=R^{2}\gamma, \qquad R^{4}=4\pi e^{\Phi_{0}}N, \label{c1}
\end{eqnarray}
with $R$ being the AdS radius. The constraint in (\ref{c1}) can be solved by introducing spherical like coordinates 
\begin{eqnarray}
\mu_{1}&=&\sin\theta\cos\psi, \nonumber\\
\mu_{2}&=&\cos\theta\label{mu i}, \\
\mu_{3}&=&\sin\theta\sin\psi,\nonumber
\end{eqnarray}
so that the metric and the KR field become
\begin{eqnarray}
ds^{2}&=&R^{2}\left\{ ds_{AdS_5}^{2}+d\theta^{2}+\sin^2\theta
  d\psi^{2}+\right. \nonumber\\
&&+G\left[\sin^2\theta\cos^{2}\psi\left(1+\hat{\gamma}^{2}\sin^2\theta\cos^{2}\theta\sin^{2}\psi\right)d\phi_{1}^{2}+\right.\nonumber\\
&&+\cos^{2}\theta\left(1+\hat{\gamma}^{2}\sin^{4}\theta\cos^{2}\psi\sin^{2}\psi\right)d\phi_{2}^{2}+\nonumber\\ 
&&+\left.\sin^2\theta\sin^{2}\psi\left(1+\hat{\gamma}^{2}\sin^2\theta\cos^{2}\theta\cos^{2}\psi\right)d\phi_{3}^{2}\right]+\nonumber\\
&&\left.+2G\hat{\gamma}^2\sin^{4}\theta\cos^{2}\theta\sin^2\psi\cos^{2}\psi\left(d\phi_{1}d\phi_{2}+d\phi_{2}d\phi_{3}+d\phi_{1}d\phi_{3}\right)\right\}, \label{LM metrica 2} \\
B_{2}&=&R^{2}\hat{\gamma}G\sin^2\theta\left(\cos^{2}\theta\cos^{2}\psi
  d\phi_{1}\wedge d\phi_{2}+\cos^{2}\theta\sin^2\psi
  d\phi_{2}\wedge d\phi_{3}-\right.\nonumber\\ &&\left.-\sin^2\theta\sin^2\psi\cos^{2}\psi d\phi_{1}\wedge d\phi_{3}\right), 
\end{eqnarray}
where now 
\begin{equation}
G^{-1}=1+\hat{\gamma}^2\sin^{2}\theta(\cos^{2}\theta+\sin^2\theta\cos^{2}\psi\sin^2\psi).
\end{equation}

\begin{figure}[h]
       \centering  
       \includegraphics[height=4cm]{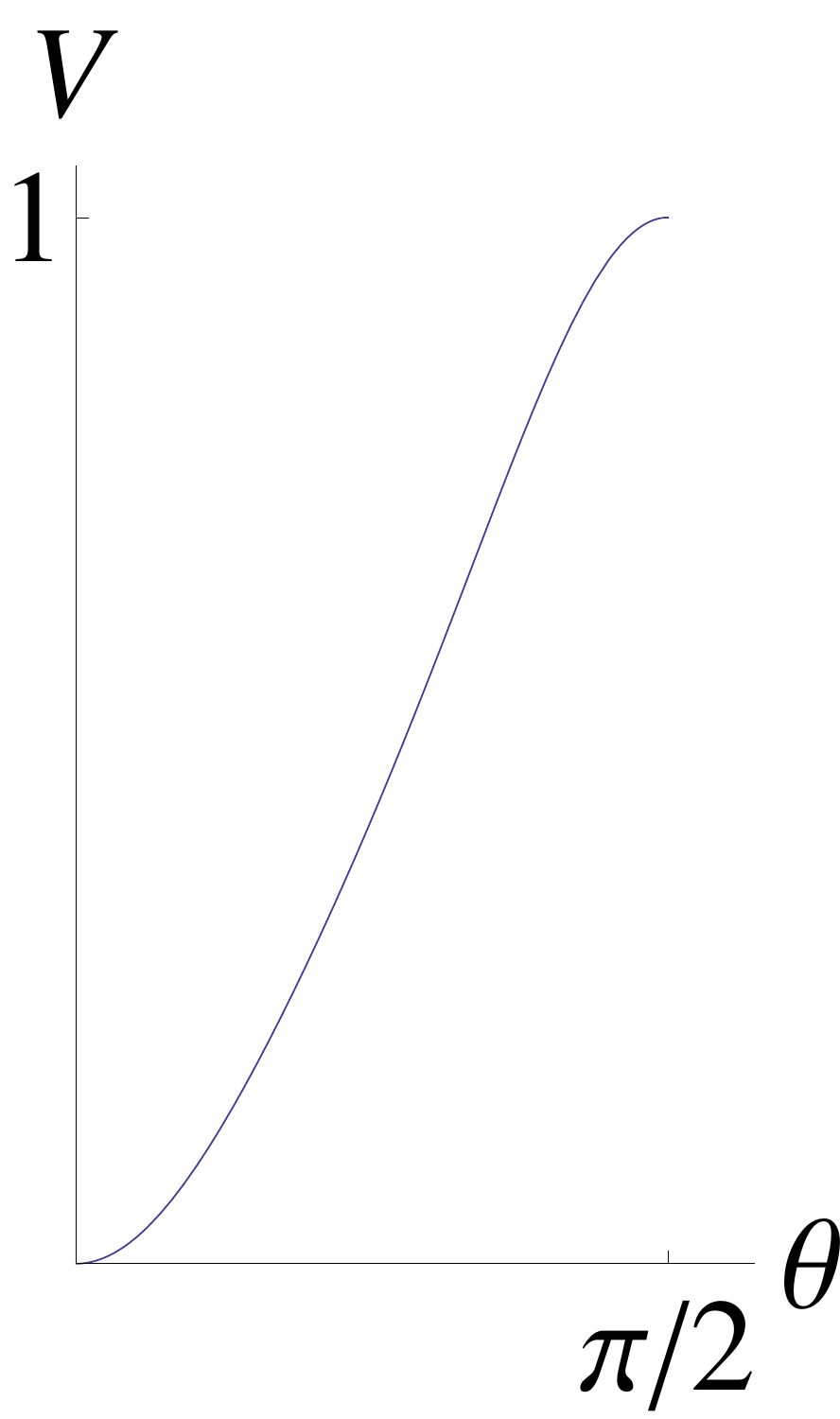}
       \caption{\it The effective potential in $AdS_5 \times S^5_{\hat{\gamma}}$}
       \label{fig4}
\end{figure}

For a pulsating string in the deformed sphere we assume that in the $AdS_5$ sector (\ref{ads_sec}) we have $\Psi=\Phi_1=\Phi_2=0$ while in the  $S^5_{\hat{\gamma}}$ sector (\ref{LM metrica 2}) we take $\phi_1=\phi_2=0$ and $\psi=\pi/2$ so that we still have a deformed sphere inside $S^5_{\hat{\gamma}}$. In this situation the relevant part of the metric reduces to 
\begin{equation} \label{metr2}
ds^{2}= R^{2}\left(-\cosh^2\rho\, dt^2+d\rho^2+d\theta^2+G\sin\theta^2\, d\phi_3^{2}\right),
\end{equation}
while $B_2=C_2=0$, and now $G = 1/(1 + \hat{\gamma}^2 \sin^2\theta \cos^2\theta)$. 
The ansatz for a pulsating string wound $m$ times along $\phi_3$ in the deformed sphere is 
\begin{equation} \label{ans2}
t = \kappa \tau,\ \qquad\rho=\rho(\tau),\qquad\theta=\theta(\tau), \qquad\phi_3=m\sigma.
\end{equation}
The equations of motion are satisfied if $\rho$ is constant, so we take the string at the center of $AdS_5$. Again we introduce an effective potential 
\begin{eqnarray}
\frac{\dot{\theta}^2}{m^2} &=& \frac{\kappa^2}{m^2} - V(\theta), \nonumber \\
V(\theta)&=& \frac{\sin^2\theta}{1+\hat{\gamma}^2\sin^2\theta\cos^2\theta},
\end{eqnarray}
which is plotted in Fig.\ref{fig4}. The potential has a maximum at $\pi/2$ with value 1 so that there are two cases to be analyzed. If $\kappa^2/m^2<1$ then there is a turning point so that $\theta$ is limited to a maximum value $\theta_+$. In this case the energy is small $E^2 < \lambda m^2$ and we have a short string oscillating on the deformed sphere. If $\kappa^2/m^2>1$ then $E^2 > \lambda m^2$ and there are no turning points. We now have a string oscillating all the way from the equator to one of the poles of the deformed sphere. 

We now use the Polyakov action for the ansatz (\ref{ans2})
\begin{equation}
	S = - \frac{\sqrt{\lambda} \kappa}{2} \int d\tau ( \cosh^2\rho - \dot{\rho}^2 - \dot{\theta}^2 + m^2 \sin^2\theta G), 
\end{equation}
to find in the case of short strings, when $\kappa^2/m^2<1$, that 
\begin{eqnarray}
	E &=& \sqrt{2 \lambda^{1/2} m N} \left[ 1 - \frac{1}{8} \frac{N}{\sqrt{\lambda} m} - \frac{5}{128} \left(\frac{N}{\sqrt{\lambda}m} \right)^2 + \dots \right] \nonumber \\
	&-& \frac{3\sqrt{2}}{8} \hat{\gamma}^2 \frac{N^{3/2}}{(\sqrt{\lambda} m)^{1/2}} \left[ 1 - \frac{41}{24} \frac{N}{\sqrt{\lambda}m} + \frac{193}{384} \left( \frac{N}{\sqrt{\lambda}m} \right)^2 + \dots \right].
\end{eqnarray}
There is a natural small expansion parameter $N/(\sqrt{\lambda} m)$ so we are in the regime of low quantum numbers. Also our results reduce to the $AdS_5 \times S^5$ results \cite{Beccaria:2010zn} in the undeformed case. We can not perform the quantization for strings with very large energy because we had to assume that $E/(\sqrt{\lambda}m)$ is small. For the case $\kappa^2/m^2>1$  we find that 
\begin{eqnarray} \label{class2}
	 E &=& N \left[ 1 + \frac{1}{4} \left( \frac{\sqrt{\lambda}m}{N} \right)^2 - \frac{1}{64} \left( \frac{\sqrt{\lambda}m}{N} \right)^4 + \dots \right] \nonumber \\
	 &-& \frac{1}{32} \hat{\gamma}^2 \lambda \frac{m^2}{N} \left[ 1 - \frac{3}{16} \left( \frac{\sqrt{\lambda}m}{N} \right)^2 + \frac{5}{128} \left( \frac{\sqrt{\lambda}m}{N} \right)^4 + \dots \right].
\end{eqnarray}
There is a small expansion parameter $\sqrt{\lambda}m/{N}$ so that we can consider the regime of large quantum numbers.

The string can be quantized using the Nambu-Goto action for the ansatz (\ref{ans2}) and we get 
 \begin{equation}
	H^2 = \cosh^2\rho \,\, [ \Pi_\rho^2 + \Pi_\theta^2 + \lambda m^2  \sin^2\theta( 1 - \hat{\gamma}^2 \sin^2\theta \cos^2\theta)] \label{HAM}.
\end{equation}
Considering the string sitting at the center of $AdS$ we will quantize it in the deformed sphere (\ref{metr2}). Then $\Pi_\theta^2$ has to be realized like (\ref{lap}) (with $r$ replaced by $\theta$) using $\sqrt{-g} = \sqrt{G} \sin\theta $ leading to 
\begin{eqnarray} \label{ham2}
	H^2 \Psi &=& - \frac{1}{\sin\theta} \frac{d}{d\theta} \left( \sin\theta \frac{d\Psi}{d\theta} \right) + \hat{\gamma}^2 \sin\theta \cos\theta ( 1 - 2 \sin^2\theta ) \frac{d\Psi}{d\theta} \nonumber \\
	&+& \lambda m^2 \sin^2\theta ( 1 - \hat{\gamma}^2 \sin^2\theta \cos^2\theta ) \Psi = E^2 \Psi.
\end{eqnarray}
If the unperturbed Hamiltonian is taken as the first term then the normalized eigenfunctions are written in terms of Legendre polynomials
\begin{equation} 
	\Psi_n(\theta) = \sqrt{2n+1} P_n(\cos\theta),
\end{equation}
For highly excited strings $E_{0,n} = n$ and it agrees with the lowest order term in (\ref{class2}) if the oscillation number is quantized as an integer. 
The first order correction to the energy is then found to be 
\begin{equation} \label{correc1}
\delta_1 E^2_n =  \frac{1}{2} \lambda m^2 \left( 1 - \frac{1}{8} \hat{\gamma}^2 \right),
\end{equation}
while the second order correction is  
\begin{equation} \label{correc2}
	\delta_2 E^2_n = \frac{1}{32} \frac{\lambda^2 m^4}{n^2} \left( 1  - \frac{1}{8} \hat{\gamma^2} \right).
\end{equation}
We then find for the energy of highly excited string states
\begin{equation} \label{fullenergy}
	E_n = n \left[ 1 + \frac{1}{4} \frac{\lambda m^2}{n^2} - \frac{1}{64} \left( \frac{\lambda m^2}{n^2} \right)^2 + \dots \right] - \frac{1}{32} \hat{\gamma}^2 \frac{\lambda m^2}{n} \left[ 1 - \frac{3}{16} \frac{\lambda m^2}{n^2} + \dots \right],
\end{equation}
in perfect agreement with the classical expression (\ref{class2}) for the energy if the oscillation number is quantized as $n$. This is expected because we are quantizing in the deformed sphere.

Effects of the full ten dimensional deformed spacetime on $\theta$ can be taking into account by considering $\sqrt{-g} = \sqrt{G} \sin^3\theta \cos\theta \dots$ in $\Pi_\theta^2$, that is, 
\begin{eqnarray}\label{4.21}
	\Pi^2_\theta &=& \frac{1}{G \sin^3\theta \cos\theta} \frac{d}{d\theta} \left( G \sin^3\theta \cos\theta \frac{d}{d\theta} \right)  \nonumber \\
	& = & \frac{1}{\sin^3\theta \cos\theta} \frac{d}{d\theta} \left( \sin^3\theta \cos\theta \frac{d}{d\theta} \right) - 2 \hat{\gamma}^2 \sin\theta \cos\theta ( 1 - 2 \sin^2\theta ) \frac{d}{d\theta}. 
\end{eqnarray}
Taking the first term of (\ref{4.21}) as the unperturbed Hamiltonian in (\ref{HAM}) the normalized eigenfunctions are expressed in terms of Jacobi polynomials as
\begin{equation}
\Psi_n(\theta) = 2 \sqrt{n+1} P_n^{(0,1)}( 1 - 2 \cos^2\theta).
\end{equation}
Then $E_{0,n} = 2n$ for large $n$ and we get agreement with the lowest order term of (\ref{class2}) if the oscillation number is quantized as an even integer. Now the wave function is even about $\theta = \pi/2$. 
The first order correction  is
\begin{equation}\label{4.23}
	\delta_1 E^2_n = \frac{1}{2} \lambda m^2 \left( 1 - \frac{1}{8} \hat{\gamma}^2 \right) - \hat{\gamma}^2, 
\end{equation}
which is precisely the result (\ref{correc1}) with an extra term which came from the one derivative contribution in (\ref{4.21}). This contribution is at the same order of $1/n$ as the other terms. In the case of the deformed Minkowski spacetime they were of different orders and the corresponding term was of lower order. 
To second order correction is 
\begin{equation}\label{4.24}
	\delta_2 E^2_n = \frac{1}{128} \frac{\lambda^2 m^4}{n^2} \left( 1 - \frac{1}{8} \hat{\gamma}^2 \right) + \frac{1}{32} \hat{\gamma}^2 \frac{\lambda m^2}{n^2},
\end{equation}
and again we find an extra term which was absent in the deformed Minkowski spacetime case. 
Notice that the Hamiltonian mixes terms of order $\hat{\gamma}^2$ and $\lambda m^2$ so that both corrections are present in the energy.
Then the energy of highly excited states is 
\begin{equation}
	E = 2n \left[ 1 + \frac{1}{4} \frac{\lambda m^2}{(2n)^2} - \frac{1}{64} \left( \frac{\lambda m^2}{(2n)^2} \right)^2 + \dots \right] - \frac{1}{32} \hat{\gamma}^2 \frac{\lambda m^2}{2n} \left( 1 - \frac{3}{16} \frac{\lambda m^2}{(2n)^2} + \dots \right) - \frac{1}{4} \frac{\hat{\gamma}^2}{n} .
\end{equation}
The extra contributions coming from the corrections to the energy in (\ref{4.23}) and (\ref{4.24}) gave rise to the last term $\hat{\gamma}^2/(4n)$. Up to this term we have agreement with (\ref{class2}) if the oscillation number is quantized as $2n$.

Finally we can incorporate some contribution from the $AdS$ sector along the lines of \cite{Minahan:2002rc}. We can consider the term $\Pi^2_\rho$ in (\ref{HAM}) and its contribution to the energy. 
We find that for the unperturbed Hamiltonian there is a shift in the energy $E \rightarrow E + 2(N_\rho+1)$ where $N_\rho$ is the quantum number associated to $\rho$. Since we are considering highly excited oscillation states we can ignore $N_\rho$ assuming that the wave function is concentrated at the origin of the $AdS$ allowing us to take $\rho=0$  so that the energy is just shifted by 2. The same result holds when we include the deformation because the structure of the equations in $AdS$ remains intact

\ack
 
Talk presented at Quantum Theory and Symmetries 7, Prague, August 7-13, 2011. 
The work of V.O.R. was supported by CNPq grant No. 304495/2007-7 and FAPESP grant No. 2008/05343-5.

\section*{References}

\bibliography{strings}

\end{document}